\documentclass[aps,pra,twocolumn,showpacs,floatfix]{revtex4}
\usepackage{graphicx}
\usepackage{color}
\usepackage{bm}
\usepackage{amsmath}
\newcommand{\eq}[1]{Eq.~(\ref{#1})}

\newcommand{\be}[1]{\begin{equation}\label{#1}}
\newcommand{\ee}{\end{equation}}

\renewcommand{\vec}[1]{{\boldsymbol #1}}

\begin{document}
\title{Attosecond light pulse induced photo-association}

\author{Paula~Rivi{\`e}re, Camilo~Ruiz and Jan-Michael~Rost}
\affiliation{Max-Planck Institute for the Physics of Complex Systems, N{\"o}thnitzer Str. 38, D-01187 Dresden, Germany.}

\begin{abstract}
We explore stimulated photo-association in the context
of attosecond pump-probe schemes of atomic matter.  
An attosecond pulse -- the probe -- is used to induce
photo-association of an electronic wave packet which had
been created before, typically with an attosecond pump pulse at an 
atomic center different from the one of photo-association.
We will show that the electron absorption is maximal for a
certain delay between the pulses.  Two ways of enhancing and controlling
stimulated photo-association are proposed, namely using an additional
infrared pulse to steer the electronic wave packet and using a
train of attosecond pulses instead of a single pair. A direct
application of ultrafast stimulated photo-association is the
measurement of atomic distances.
\end{abstract}


\maketitle

\section{Introduction}
  Attosecond laser pulses hold the promise to be a tool to image structure of
  matter on the atomic scale with unprecedented temporal and spatial
  resolution \cite{Paul2001,LopezMartens2005,Corkum2007}.  Nowadays,
  attosecond pulses with a typical duration of electronic motion in
  atoms or molecules \cite{Kienberger2002} can be produced and
  characterized in a controlled way \cite{Nabekawa2006}, and are
  therefore ready to be used for mapping and controlling electronic
  motion \cite{Kling2006,Feng2007}.  First applications of attosecond
  pulses have focused on the generation of attosecond electron
  wavepackets, e.g., for real time observation of tunneling
  \cite{Uiberacker2007}, the interference of wavepackets using
  attosecond trains \cite{Lund2007}, or to obtain autoionization widths
  \cite{Drescher2002}. They have also been used for imaging molecular
  orbitals \cite{Itatani2004}.
All the applications show that the role of attosecond pulses has been
restricted so far to study ionization or interference processes.  Here
we propose a new application of attosecond pulses: the recapture of a
continuum electron by an ion, induced by an attosecond pulse.  This
constitutes a new process of photo-association, namely induced
photo-association (IPA).  Although it is a process of stimulated photo-absorption
triggered by the attosecond photon field, it has low
probability due to the small timespan of the attosecond pulse during
which the IPA is possible.

Hence, after the demonstration of the basic process of IPA we will
also discuss how the reabsorption probability can be enhanced.  This
can be achieved by using an attosecond pulse train, which delivers a
series of pairs of pump-probe pulses.  The drawback of such a
technique is that the time delay between two attosecond pulses is
typically not variable in an experiment.  However, adding an infrared
(IR) pulse can achieve in the context of IPA the same effect as
changing the delay between the attosecond pulses, namely controlling
the time when the electronic wavepacket arrives at the atomic center
by which it is supposed to be captured.  This will be shown
explicitely for one attosecond pulse per IR cycle, which is
experimentally achievable \cite{Mauritsson2006}.

 Since this work is mainly intended to introduce the concept of IPA,
 we restrict ourselves to the simplest possible scenario, two ions and
 an electron, and to be specific, we take  $H_{2}^{+}$. Furthermore,
 we carry out fully time-dependent quantum calculations in one and two
 spatial dimensions. 

 The paper is organized as follows. In section II we introduce our
 Hamiltonian and describe how we solve the electron dynamics. In
 section III we describe the basic IPA process with two attosecond
 pulses. In section IV we discuss how the addition of an IR pulse
 changes the dynamics. Section V presents the results for an
 attosecond pulse train in combination with an IR pulse, and section
 VI contains conclusions and outlook of the paper.
 Atomic units are used unless stated otherwise.

 \section{Time-dependent  one-electron quantum dynamics in a landscape
of attractive potentials}

As mentioned before, for the sake of clarity of the concept, we
consider the simplest possible dynamical situation for IPA to occur in
a pump-probe setting with the ions (protons) fixed in space.  We have
in mind the idealized situation that an electron, bound to one proton
(forming hydrogen) is ionized by the first atto pulse.  The electron
packet propagates and part of it is recaptured subsequently at another
proton, assisted by a second atto pulse.  The recapture is maximal if
the probe pulse comes at the time when the center of the wavepacket is
at the second proton.

The fixed nuclei assumption is a twofold but well justified
approximation. Firstly, the situation of a hydrogen atom and a proton
separated by 40 a.u. as considered below, is naturally created by
exciting the $H_2^+$ molecule from its ground state, with equilibrium internuclear distance $R_0=2$, to a dissociative state. 
This leads to a maximum nuclear kinetic energy
determined by the energy of this state, which is $V_{2p\sigma_u}(R_0) = -0.169$.
At a distance $R=40$, the potential energy approaches $V_{2p\sigma_u}(\infty)=-0.5$. The difference  
determines the maximum kinetic energy, $Mv^2=V_{2p\sigma_u}(R_0)-V_{2p\sigma_u}(\infty)$. 
Hence the relative velocity of the two nuclei with mass $M=1836$ is $v\sim 0.0134$,
so that the nuclei move less than half an atomic
unit during a typical pump-probe interval of about 50 a.u..  Even
smaller is the motion due to the IPA process itself, i.e., due to the
Coulomb repulsion of the two protons after the first (ionizing) 
attosecond pulse.

We assume the light field to be linearly polarized along the x-axis (unit vector
$\hat x$),
\begin{equation}
\vec{E}(t)=\hat{x}E_{0}\left[\sin(\omega t)\sin^2\left(\frac{\omega t}{2N}\right)
+\sin(\omega t')\sin^2\left(\frac{\omega t'}{2N}\right)\right]\,,
\label{field}
\end{equation}
where $t'=t-\Delta t$ and $\Delta t$ is the delay between the two
attosecond pulses, which have a central frequency $\omega$ and a duration of $N$ cycles.
The Hamiltonian reads
\begin{eqnarray}
    H = \frac{\vec p^{2}}{2}+V+p_{x}A(t)/c \equiv H_{0}+ p_{x}A(t)/c\,, \label{ham}
\end{eqnarray}
where $A(t)=-c\int^t E(t')\,dt'$.
Furthermore, we consider only the case where the two protons are 
fixed on the x-axis at distances $\pm x_{0}$ with $x_{0}=20$,
so that the soft-core potential reads in two dimension $V(x,y)=
W_{-}(x,y)+W_{+}(x,y)$ with
\begin{equation}
W_{\pm}(x,y)=-[(x\mp
x_{0})^{2}+y^{2}+\epsilon]^{-\frac 12}\,. 
\label{pot}
\end{equation}
The softening parameter $\epsilon$, adjusted to provide a ground state
energy of $E_{g}= -0.5$ for the isolated $H$ atom, is
$\epsilon_{1D}=2.0$ and $\epsilon_{2D}=0.63$ in one and two
dimensions, respectively.
The ground state energy corresponds to
an isolated hydrogen atom, which is approximately true for the second
proton at a distance of 40 atomic units. To localize the state at one nucleus,
one has to take superpositions of the ground states of gerade and
ungerade symmetry, $\psi_{g/u}$, which are formally the ground and
first excited state of the system. This is well known from ion-atom
collisions,
\begin{equation}
\psi_\pm=(\psi_{g}\pm\psi_{u})/\sqrt{2},
\label{initialstate}
\end{equation}
where we define $\psi_{+}$ to be localized at the right well
$W_{+}(x,y)$ at $x_{0}=+20$  of \eq{pot} \footnote{Note that in three dimensions, $\psi_{g/u}$ are
given by the ground state $1s\sigma_{g}$ and first excited state
$2p\sigma_{u}$ molecular orbitals.}.

The initial state is propagated numerically in time under the
Hamiltonian of \eq{ham} using the Crank-Nicholson method with a time
step of $\delta t=0.05$.  The grid ranges from $-100$ to $+100$, with a
spatial step $\delta x=0.1$.  The eigenstates $\psi_{g/u}$ from
\eq{initialstate} are computed using imaginary time-propagation under
the Hamiltonian $H_{0}$ from \eq{ham}.

\section{Two attosecond pulses}\label{section1}
In this section we will investigate under which conditions induced
photo-association can be realized within the framework of a dual
attosecond pulse pump-probe scenario \cite{Hu2006}.  For simplicity,
we will assume that at the beginning the electron is always well
localized around $x_{0}$ in the well $W_{+}$ with initial wavefunction
$\psi_{+}(t=0)$.  Asymmetric dissociative states can be realized (with 
an electron localization probability up to 84$\%$) by
means of laser-assisted dissociation \cite{Feng2007}.  We use three
different wavelengths (40 nm, 60 nm and 80 nm), and high intensity
attosecond pulses ($I = 10 ^{16} $W/cm$^{2}$) in order to obtain good
statistics for IPA.
\begin{figure}
\includegraphics[width=1.1\columnwidth]{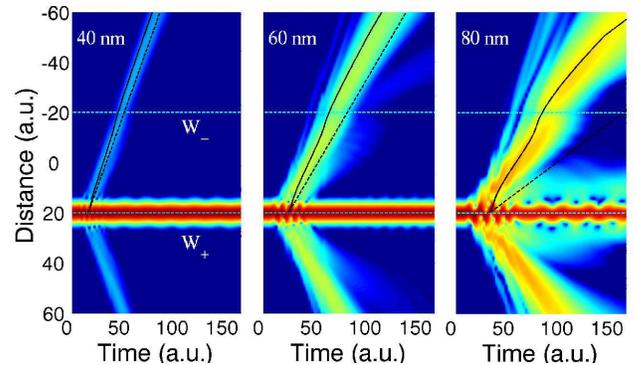}
\caption{Electron density as a function of time, under the action of a single
attosecond pulse of 40, 60 and 80 nm central wavelength and $N=6$ cycles
(the central time of the pulse is approximately 16.5, 25 and 33 a.u., respectively).
The wells with potentials $W_\pm$ are located at $\pm 20$ a.u., marked by horizontal 
dashed lines. The electronic ionization path predicted according to 
\eq{velocity} (solid) and \eq{velocity2} (dashed) is indicated.}
\label{fig1}
\end{figure}

For optimum recombination, the main velocity $v$ of the electron
wavepacket, the delay $\Delta t$ between the pulses and the distance
$2x_{0}$ between the two atomic centers must fulfill roughly the
classical relation $v =2x_{0}/\Delta t$.  More precisely, since the
electron absorbs one photon of energy $\omega$ from the pump pulse,
energy conservation can be used to determine the electron velocity
of the center of the wavepacket classically for a calculated electron
trajectory $\vec r(t)$. With the Hamiltonian $H$  from \eq{ham}
we have, for $t>2\pi N/\omega$,
\begin{equation}
\frac{p^{2}(t)}{2}+V(\vec r(t)) = \omega + E_{i}.
\label{energy}
\end{equation}
Hence,
\begin{equation}
v(t) = p(t) = \left\{2[\omega + E_{i} - V(\vec r(t))]\right\}^{1/2},
\label{velocity}
\end{equation}
and ignoring the potential energy for electrons with large energy in
the continuum, the velocity becomes time-independent,
\begin{equation}
v(t) \approx v^{(0)} = \left[2(\omega + E_{i})\right]^{1/2}.
\label{velocity2}
\end{equation}
Results for the action of one attosecond pulse with $\lambda=$40 nm
($\hbar \omega=$1.14), 60 nm (0.76) and 80 nm (0.57) are shown in
Fig.~\ref{fig1} for $E_{i}=E_{g}$.  The attosecond pulse has a
duration of $N=6$ cycles, and an intensity $I=10^{16}$~W/cm$^2$.  The
classically calculated path of the electron is shown for both, free
electronic motion (dashed lines, \eq{velocity2}) and taking into
account the nuclear potential (full lines, \eq{velocity}).  While the
free electron approximation is reasonably good for the most energetic
electrons (40 nm) with excess energy of $\omega+E_g = 0.614$ (at a
wavelength of 40 nm), it is not valid any more for the lower excess energy
of 0.045 (80 nm).

The time evolution of the eigenstates under the influence of the total
Hamiltonian given by \eq{ham} is
\begin{equation}
\psi_{\pm}(\vec r,t)=U(t)\psi_{\pm}(\vec r,0),
\end{equation}
where $U(t)$ is the time-evolution operator. We can also define the 
time evolution of the eigenstates under $H_0$, which is time-independent,
\begin{equation}
\psi_{\pm}(\vec r,t)=U_0(t)\psi_{\pm}(\vec 
r,0)\equiv\exp^{-iH_0t}\psi_{\pm}(\vec r,0).
\end{equation}
As a measure for the fraction of the electron wavepacket which gets recaptured at the left 
well $W_-$, we calculate the overlap of its eigenfunction $\psi_-(x,t)$
(see \eq{initialstate}) with the time-dependent wavepacket
\begin{equation}
C_{-}(t)=\langle \psi_-(\vec r,t) |\psi(\vec r,t)\rangle,
\end{equation}
which is formally a cross-correlation function.

\begin{figure}
\includegraphics[width=0.85\columnwidth]{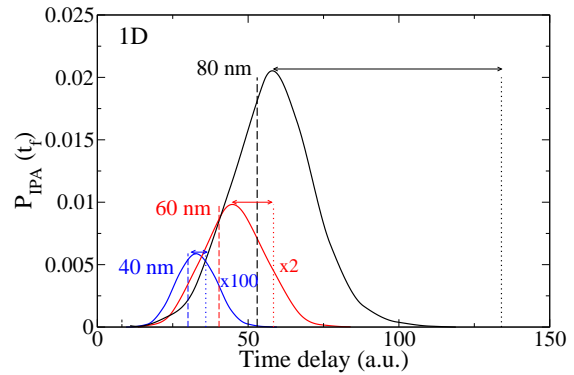}
\caption{
Probability of the induced
photo-association at $W_-$ as a function of the delay between the
pulses, for three photon wavelengths: $\lambda=$40, 60 and 80 nm
(one-dimensional calculations).  The predicted optimal delay according
to \eq{velocity} and \eq{velocity2} is indicated with dashed and
dotted lines, respectively. The arrows highlight the difference between
the real and calculated values if the potential energy is ignored.
}
\label{fig3}
\end{figure}

The probability for induced photo-association (IPA) in the left well 
is therefore given by
\be{IPA}
P_\mathrm{IPA}(t)=|C_-(t)|^2\,,
\end{equation}
The final population of $W_-$ is 
$P_\mathrm{IPA}(t_f)$, 
where $t_f$ is the time at the end of the second attosecond pulse,
which is delayed by $\Delta t$ with respect to the first one.
As can be seen from Fig.~\ref{fig3}, an optimum delay $\Delta t^*$ for maximum IPA exists.
$\Delta t^*$ is reasonably predicted by the high photon energy
approximation for 40 nm, but not for lower energies (dotted lines).
However, the inclusion of the nuclear potential (\eq{velocity}, dashed
lines) provides a good prediction for all three energies, with
differences with respect to the real value of around 8$\%$ for 40 nm,
and 10$\%$ for 60 and 80 nm.  Note, that larger wavelengths provide a
worse accuracy in the predicted optimum delays, but correspond to
higher absorption probabilities, which might be useful experimentally.

The experimental observation of IPA is possible by detecting the
outgoing proton or electron flux.  For optimal time delay sufficient
absorption at $W_-$ will lead to a decrease in the number of protons
as well as the number of electrons detected in this direction
(negative x-axis). 

\begin{figure}
\includegraphics[width=0.85\columnwidth]{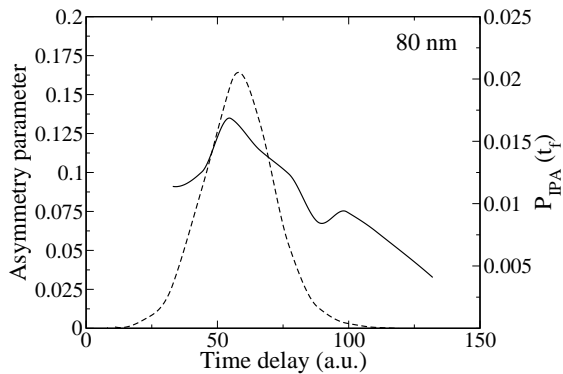}
\caption{Full line: Asymmetry parameter as a function of the time delay (\eq{asimeq}), 
for $\lambda=80$ nm and the same laser parameters as in Fig.~\ref{fig3}.
Dashed line: IPA probability (population at $W_-$), shown for comparison.}
\label{asim}
\end{figure}

To study the asymmetry in the electronic emission, we use the
continuity equation
\begin{equation}
\frac{\partial|\Psi(\vec{r},t)|^2}{\partial t}=-\vec{\nabla}\vec{{j}}(\vec{r},t),
\label{conteq}
\end{equation}
which relates the change in the electronic density as a function of
time with the particle flux
\begin{equation}
\vec{{j}}(\vec{r},t)=\Re\left(\Psi^*(\vec{r},t)\frac{\hbar}{im}\vec{\nabla}\Psi(\vec{r},t) \right)\,.
\end{equation}
In one dimension, integration  of \eq{conteq} in space leads to
\begin{equation}
\int_{-L}^{L}\frac{\partial |\Psi(x,t)|^2}{\partial t}\,dx \equiv 
\frac{\partial N}{\partial t} = 
j(-L,t)-j(L,t)\,,
\label{intcont}
\end{equation}
where $N$ is the number of electrons. Using \eq{conteq} again for the 
right hand side of  \eq{intcont} yields
\begin{equation}
\frac{\partial N}{\partial t}= \frac{\partial N_{+}}{\partial 
t}+\frac{\partial N_{-}}{\partial t}\,,
\end{equation}
providing the ionization rate $\frac{\partial N}{\partial t}$ in
terms of  particles $N_{+}$ escaping to the right passing the position $+L$ and 
the ones $N_{-}$  escaping to the left passing $-L$.
 We define the parameter
\begin{equation}
A=\frac{\Delta N_+-\Delta N_-}{\Delta N_++\Delta N_-}
\label{asimeq}
\end{equation}
for asymmetric escape.
If the electron is originally located in the well $W_+$ (at $+r_0$), the flux
should ideally be symmetric if there is no absorption at $W_-$.
With absorption at $W_-$  located at $-r_0$, there should be less flux
through $-L$ than through $L$, so that $A>0$. 
 In Fig.~\ref{asim} the asymmetry parameter for the two wells system
 is shown (full line) together with the IPA probability (dashed line,
 see also Fig.~\ref{fig3}) for comparison.  Indeed, $A$ has a maximum
 if the population at the second well is maximal.

Note, that the asymmetry parameter is never zero. 
The second well reflects part of the ionized wavefunction, resulting
in a higher value of ionization flux towards the side where $W_+$ is
located (positive x-axis).  For this reason the experimental detection
of the reabsorption in $W_-$ should be based on the shape of the
asymmetry parameter, and not on its absolute value.

Our one-dimensional calculations suggest that electron population can
be efficiently transferred to the second well $W_-$ by optimizing the
time delay between two attosecond pulses.  The induced photo
association probability as a function of the time delay can be used to
measure the distance of the nuclei in the molecule.  Furthermore, we
have seen that the left and right asymmetry in the ionization signal
could be a natural way to observe these effects in an experiment.

\begin{figure}
\includegraphics[width=0.85\columnwidth]{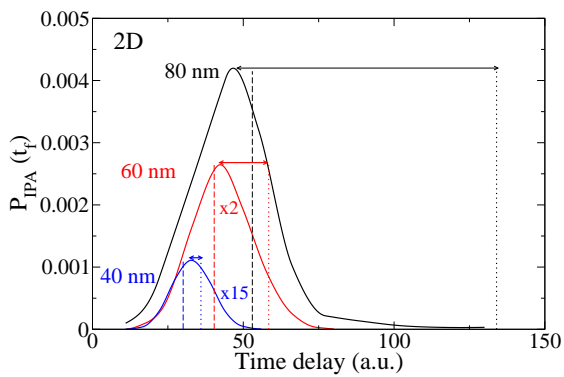}
\caption{The same as Fig.~\ref{fig3} but in two dimensions.
}
\label{proy2D}
\end{figure}

To substantiate these findings we have carried out more realistic 2D calculations,
where spreading of the electronic wavepacket in the 
coordinate transversal to the laser polarization is naturally included.
Results shown in Fig.~\ref{proy2D} are fully analogous to those of the
one-dimensional case, although the IPA probabilities are obviously
smaller, since the outgoing electron has now access to the transverse
direction.  The differences between obtained and predicted maxima
according to \eq{velocity} are now 12$\%$ for 80~nm, 6$\%$ for 60~nm
and 9$\%$ for 40~nm.
\section{Two attosecond pulses plus Infrared field}
So far we have demonstrated that the time delay between two attosecond
pulses can be used to measure atomic distances, or to optimize the
electronic transfer between two nuclei.  However, with the current
state-of-the-art technology, it is still difficult to isolate two pulses
and choose the time delay between them.

Attosecond pulses are created \cite{Paul2001} through High Harmonic
Generation produced by the interaction of an IR laser field with an
atom or molecule.  With this technique one can easily generate
Attosecond Pulse Trains (APT), with one or two \cite{Mauritsson2006}
attosecond pulses per infrared cycle of the generating field.  For a
Ti:Sa laser ($\lambda = $800 nm), the time delay between attosecond
pulses is fixed to $T\sim$110 or $T/2\sim$55, and therefore the 
technique described in the previous section cannot be directly applied.
\begin{figure}
\includegraphics[width=0.85\columnwidth]{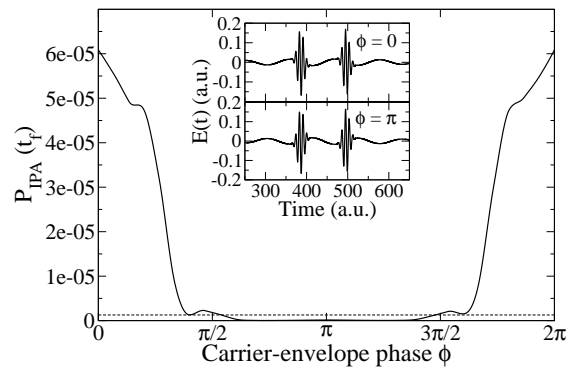}
\caption{Probability of the induced photo-association at $W_-$ as a
function of the carrier-envelope phase of the infrared pulse.  The
insets show the total field for $\phi=0$ and $\pi$.  }
\label{figair}
\end{figure}

To overcome this limitation we will introduce, in addition to a fixed
and realistic delay between attosecond pulses, a synchronized IR
field.  Due to the way the APT is generated, a synchronized IR pulse
is always available.

The phase of the IR laser pulse can be chosen to either accelerate or
decelerate the electron when it is released at the first well $W_+$
\cite{Johnsson2005,Remacle2007} and therefore it will influence the temporal
and spatial overlap between the ionized electron and the second well
$W_-$ at the time of the second attosecond pulse.

In other words, an appropriate infrared field, applied while
the electron travels the distance between the nuclei, can alter the
electron trajectory such that the electron arrives at $W_-$ when the
second attosecond pulse is applied, thereby optimizing IPA.

Our electric field has from now on three contributions: two Gaussian
attosecond pulses of intensity $I_{UV}=10^{15}$ W/cm$^2$ and $FWHM=2T$, and
an infrared field
\begin{equation}
E_\mathrm{IR}(t)=E_\mathrm{IR}\sin(\omega_\mathrm{IR}t+\phi)
\sin^2\left(\frac{\omega_{IR} t}{2N_\mathrm{IR}}\right)
\label{fieldir}
\end{equation}
with $N_{IR}=8$ and $I_{IR}=10^{13}$ W/cm$^2$.  We have reduced
the UV intensity by a factor of 10 compared to the previous section, to
work out more clearly the IPA enhancement produced by the IR field.  The
two attosecond pulses, separated by an infrared period $\Delta
t=T_{IR}$, are located symmetrically with respect to the center of the
infrared.  The central wavelength of the attosecond pulses is
$\lambda=65$ nm ($\omega=$0.7).  This implies that the optimal delay
(following \eq{velocity2}) would be $t=2x_0/v\sim 68$, much less than
the actual delay of $T=110$ between the pulses.  Therefore, we expect
enhanced recombination at $W_-$ if the IR field decelerates the
outgoing electron.
To this end we vary the carrier-envelope phase  of the infrared
pulse, $\phi$ in \eq{fieldir}.  This changes the value of the IR field
at the moment when the electron is ionized by the first attosecond
pulse and is equivalent to changing the time at which the attosecond
pulses occur during the infrared pulse.  The latter has already been
achieved experimentally \cite{Johnsson2007}.

For the present intensity, the infrared pulse by itself causes
negligible absorption at $W_-$ of only
$P_\mathrm{IPA}\sim$10$^{-13}$.  This ensures that IPA comes from the
attosecond pulse only, while the infrared pulse has exclusively the
role of accelerating or decelerating the ionized electrons.
\begin{figure}
\includegraphics[width=1.1\columnwidth]{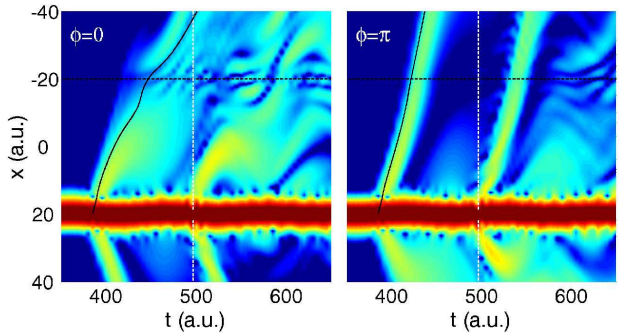}
\caption{Bending of the electron trajectory caused by the infrared
field ($\lambda=$800 nm, $I=10^{13}$ W/cm$^2$).  Left: with
carrier-envelope phase $\phi=0$ in the infrared pulse (see insets in
Fig.~\ref{figair}).  Right: the same situation when $\phi=\pi$.
Dashed black line: position of the nucleus $W_-$.  Dashed white line:
position of the center of the second attosecond pulse.  Full line:
expected trajectory following \eq{velocity}.}
\label{fig2air}
\end{figure}
Results for IPA in the second well as a
function of the carrier-envelope phase of the infrared pulse are shown
in Fig.~\ref{figair}.  The insets show the position of the attosecond
pulses with respect to the infrared pulse, for the cases with $\phi=0$
and $\phi=\pi$.  The dashed line close to the bottom of the figure
($P_\mathrm{IPA}\sim$1.275$\times$10$^{-6}$) corresponds to the case
in which only the attosecond pulses, but not the infrared, are acting
on the system.  It illustrates impressively the enormous capability of
the infrared pulse to steer the ionized electron, thereby dramatically
changing IPA: for $\phi=0$ the photo-association probability is
$\sim$50 higher than without the infrared, while for $\phi= 0.73\pi$
it is around 37 times smaller.  The contrast between maximum
($\phi=0$) and minimum ($\phi=\pi$) is therefore $\sim$1850 rendering
this mechanism very efficient.
Fig.~\ref{fig2air} illustrates the effect of the IR field on
attosecond induced photo-association.  It shows the change in the
electron trajectory due to the infrared field, for $\phi=0$ and $\pi$,
together with the expected trajectory of the electron under the
influence of the two nuclei and the IR field.  These trajectories have
been calculated with \eq{velocity}, where now
$V_\mathrm{IR}(x(t))=V+xE_{IR}(t)$, i.e., the interaction is given by
the original potential $V$ from \eq{ham} with the IR-field added.

\begin{figure}
\includegraphics[width=0.9\columnwidth]{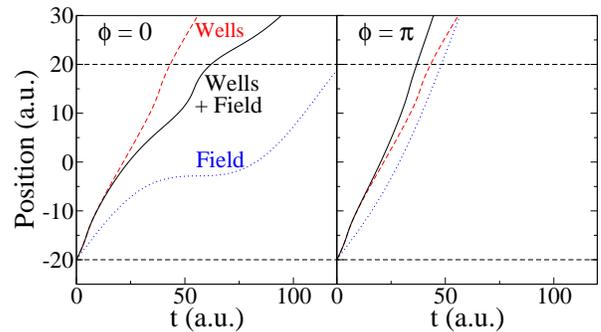}
\caption{Electron trajectories calculated with \eq{velocity} in the
presence of the wells (dashed lines), the infrared field (dotted lines),
or both of them (full lines), for carrier-envelope phases of $\phi=0$
or $\phi=\pi$.  The position of the wells is indicated by the dashed
lines.  The origin in time is the center of the first attosecond
pulse.}
\label{fig3air}
\end{figure}

With $\phi=0$, the electron is maximally decelerated.  It arrives
later at the second well, enhancing the recombination probability.
For $\phi=\pi$ the effect is a slight acceleration: the electron
arrives even earlier at the second well, hence the IPA probability is
vanishing small.  This can be seen even more clearly in
Fig.~\ref{fig3air} in terms of classical trajectories of the electron
for the cases shown in Fig.~\ref{fig2air}.  From there it is clear,
that the optimum situation for IPA (coincidence in space and time of
electron trajectory, second well position and peak of second
attosecond pulse) cannot be reached for the present parameters chosen,
which was done on purpose to provide a ``typical'' situation.
\section{Train of attosecond pulses }
The combination of two attosecond pulses and an IR pulse as discussed
in the previous section can enhance the IPA probability.  Yet, it
is still small, especially considering that in one dimension the
recombination probabilities are unrealistically increased due to the
absence of wavepacket spreading in the transversal direction.  One may
overcome this problem by using more than two
attosecond pulses.  With more pulses in the train and a longer IR
pulse, the transfer process is repeated each cycle, thus increasing
the net probability of recombination at $W_-$.  From the experimental
point of view, longer APT are even easier to generate and to handle.

After the second pulse in the train, the population at the second well
is non-zero.  Hence, for any further pulse, there will be
ionization from both wells.  Yet, given a fixed ionization
probability $p$, there is a net increase of population at $W_-$,
because initially the probability at $W_+$ is much larger than at
$W_-$.

\begin{figure}
\includegraphics[width=0.85\columnwidth]{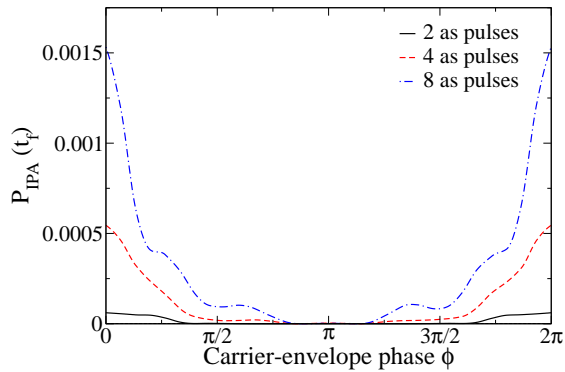}
\caption{Probability of induced photo-association at $W_-$ as a function of the carrier-envelope phase of the infrared pulse, for 2, 4 and 8 attosecond
pulses in the train. Intensities: $I_{IR}=10^{13}$ W/cm$^2$, and $I_{UV}=10^{15}$ W/cm$^2$.}
\label{figtrain}
\end{figure}

For $n$ pulses, a simple calculation of the population probabilities 
at $W_+$ and $W_-$ yields
\begin{eqnarray}
\label{popu}
A^{+}_n&=&A^{+}_{n-1}(1-p)+A^{-}_{n-2}p^2\nonumber\\
A^{-}_n&=&A^{-}_{n-1}(1-p)+A^{+}_{n-2}p^2\,.
\end{eqnarray}
 If initially the first
well is occupied ($A^{+}_0=1$) and the second well is empty ($A^{-}_0=0$), the
population of the second well as a function of the number of pulses
$n$ will evolve to a maximum whose position depends on the value of
$p$.  In our case, the first attosecond pulse induces an ionization of
about 3.64$\%$ of the initial state, thus $p=0.0364$ and the maximum
appears at $n=29$.  
In Fig.~\ref{figtrain} we show the IPA in $W_-$ versus the
carrier-envelope phase for numerical calculation using an APT
containing 2, 4 or 8 pulses.  By increasing the number of
attosecond pulses, the absorption can be better controlled: for 2 pulses
the enhancement factor at $\phi=0$ with respect to the case without IR
is 50 (as seen in Fig.~\ref{fig3}) increasing to 480 and 1530 for 4  
and 8 pulses, respectively. 

\section{Conclusions and Outlook}
We have  introduced the process of 
attosecond pulse induced photo-association.
For that purpose, a pump-probe scheme of attosecond pulses has been
used in a system consisting on a dissociating $H_2^+$ molecule with
the electron initially located in one of the ions.  The pump pulse
ionizes the electron and the probe pulse can induce the reabsorption
at the other ion when the delay between the pulses corresponds to the
time the electron needs to reach the second ion.

We have proven that the IPA probability can be controlled by changing
the delay between the pulses.  We have also discussed and analyzed two
experimental feasible ways for enhancing this probability: the use of
an infrared pulse to accelerate or decelerate the electron, and the
use of a train of attosecond pulses instead of a single pair, to
maximize the charge transfer between them.  Both ways are shown to
enhance IPA considerably.  Furthermore, the experimental detection of
reabsorption is facilitated through a study of the forward/backward
asymmetry of the electronic flux along the polarization direction.

Deliberately, we have chosen the laser parameters and the distances of,
as well as the binding energies at the atomic centers such that the
resulting dynamics can be described quasiclassically, providing an
intuitive picture in terms of classical trajectories guiding the
electronic wavepackets.  This has been important to allow for a
straightforward interpretation of the IPA results, as presented here.
In fact, IPA dynamics is far more intricate, especially, outside the
quasi-classical regime, e.g., if the excess energy of the electronic
wavepacket is small (due to a small frequency of the attosecond pulses
or a large binding energy, as studied for an atom in \cite{Johnsson2007}). 
Future work on more than one atomic center will engage in this direction
with the chance to uncover so far unknown effects.

Moreover, our ideas can be extended to more complex molecules, where
several centers may exist, opening new possibilities to carry out
microscopic time-of-flight measurements to explore microscopic spatial
landscapes.  This will be especially useful in situations where the
static and well developed diffraction methods fail due to a fast
change of the microscopic landscape, e.g., a fast explosion or
reorganization of the ions in a large molecule.
\acknowledgments
We acknowledge contributions by S. Baier, A. Becker, F. He, P. Panek
and A. Requate to the virtual laser lab Nonlinear Processes in Strong 
Fields Library, which has been used for the present calculations. 
P. Rivi{\`e}re acknowledges a postdoctoral fellowship of the
Ministerio de Educaci{\'o}n y Ciencia (Spain).


\begin{thebibliography}{99}
\bibitem{Paul2001}
P. M. Paul {\it et al.}
Science {\bf 292}, 1689 (2001)
%
\bibitem{LopezMartens2005}
R. L{\'o}pez-Martens, \emph{et al.}, 
Phys. Rev. Lett. {\bf 94}, 033001 (2005) 
%
\bibitem{Corkum2007}
P. Corkum \emph{et al.}, Nature Phys. \textbf{3}, 381 (2007)
%
\bibitem{Kienberger2002}
R. Kienberger, M. Hentschel \emph{et al.} Science {\bf 297}, 1144 (2002)
%
\bibitem{Nabekawa2006}
Y. Nabekawa \emph{et al.}, 
Phys. Rev. Lett. {\bf 96}, 083901 (2006) 
%
\bibitem{Kling2006}
M. F. Kling \emph{et al.}, Science \textbf{312}, 246 (2006)
\bibitem{Feng2007}
F. He, C. Ruiz and A. Becker, Phys. Rev. Lett. \textbf{99}, 083002 (2007)
\bibitem{Uiberacker2007}
M. Uiberacker \emph{et al.}, Nature \textbf{446}, 627 (2007)
\bibitem{Lund2007}
T. Remetter, P. Johnsson, J. Mauritsson \emph{et al.}, Nature Phys. \textbf{2}, 323 (2006)
%
\bibitem{Drescher2002}
M. Drescher \emph{et al.}, Nature \textbf{419}, 803 (2002)
%
\bibitem{Itatani2004}
J. Itatani, J. Levesque, D. Zeidler \emph{et al.}, Nature \textbf{432}, 867 (2004)
%
\bibitem{Mauritsson2006}
J. Mauritsson, P. Johnsson, E. Gustafsson \emph{et al.}, 
Phys. Rev. Lett. {\bf 97}, 013001 (2006) 
%
\bibitem{Hu2006}
S. X. Hu and L. A. Collins,
Phys. Rev. Lett. {\bf 96}, 073004 (2006) 
%
\bibitem{Johnsson2005}
P. Johnsson, R. L{\'o}pez-Martens, S. Kazamias \emph{et al.},
Phys. Rev. Lett. {\bf 95}, 013001 (2005)
%
\bibitem{Remacle2007}
F. Remacle, M. Nest and R. D. Levine,
Phys. Rev. Lett. {\bf 99}, 183902 (2007) 
%
\bibitem{Johnsson2007}
P. Johnsson {\it et al.}, Phys. Rev. Lett. {\bf 99}, 233001 (2007)
%
\bibitem{Williams2007}
I. D. Williams \emph{et al.}, 
Phys. Rev. Lett. {\bf 99}, 173002 (2007) 
%
\bibitem{MiajaAvila2006}
L. Miaja-Avila \emph{et al.},
Phys. Rev. Lett. {\bf 97}, 113604 (2006) 
%
\end{thebibliography}
\end{document}